\documentclass[runningheads]{llncs}
\usepackage{caption}
\usepackage{subcaption}

\usepackage{graphicx}
\usepackage{amsfonts}       
\usepackage{mathrsfs}
\usepackage{amsmath,amssymb}
\usepackage{tikz}
\usepackage{soul}
\usepackage{breakcites}
\usepackage[final]{pdfpages}
\usepackage[normalem]{ulem}

\let\llncssubparagraph\subparagraph
\let\subparagraph\paragraph

\usepackage{titlesec}
\let\subparagraph\llncssubparagraph
\titlespacing\section{0pt}{4pt plus 0pt minus 2pt}{4pt plus 0pt minus 2pt}
\titlespacing\subsection{0pt}{4pt plus 0pt minus 2pt}{4pt plus 0pt minus 2pt}
\titlespacing\subsubsection{0pt}{4pt plus 0pt minus 2pt}{4pt plus 0pt minus 2pt}
\usepackage{hyperref}

\begin{document}

\title{Self-supervised Lesion Change Detection and Localisation in Longitudinal Multiple Sclerosis Brain Imaging \thanks{This paper was partially supported by an Avant Doctor in Training Research Scholarship and the Australian Research Council through grants DP180103232 and FT190100525.}}

\titlerunning{Change Detection in Longitudinal Multiple Sclerosis Brain Imaging}

%
\author{Minh-Son To\inst{1,2} \and 
Ian G Sarno \inst{2} \and
Chee Chong \inst{2,3} \and
Mark Jenkinson \inst{4,5} \and
Gustavo Carneiro \inst{5}}

\authorrunning{M. To et al.}



%

%

\institute{FHMRI, Flinders University, Australia \email{minhson.to@flinders.edu.au}  \and
SAMI, Flinders Medical Centre, Australia \and
Dr Jones \& Partners Medical Imaging, Australia \and
FMRIB Centre, University of Oxford, United Kingdom \and
AIML, University of Adelaide, Australia}


%
\maketitle              
\begin{abstract}

Longitudinal imaging forms an essential component in the management and follow-up of many medical conditions. The presence of lesion changes on serial imaging can have significant impact on clinical decision making, highlighting the important role for automated change detection. 
Lesion changes can represent anomalies in serial imaging, which implies a limited availability of annotations and a wide variety of possible changes that need to be considered. Hence, we introduce a new unsupervised anomaly detection and localisation method trained exclusively with serial images that do not contain any lesion changes. Our training automatically synthesises lesion changes in serial images, introducing detection and localisation pseudo-labels that are used to self-supervise the training of our model. Given the rarity of these lesion changes in the synthesised images, we train the model with the imbalance robust focal Tversky loss. When compared to supervised models trained on different datasets, our method shows competitive performance in the detection and localisation of new demyelinating lesions on longitudinal magnetic resonance imaging in multiple sclerosis patients. Code for the models will be made available on GitHub.


\end{abstract}
\section{Introduction}

Diagnostic imaging interpretation routinely involves comparisons with prior imaging to identify new lesions or detect changes to existing structures and lesions.
Change detection can be a tedious and difficult task for the radiologist~\cite{Altay2013} and importantly, the presence of change can alter clinical management. 
For example, magnetic resonance imaging (MRI) plays a central role in monitoring disease progression in multiple sclerosis~\cite{Rovira2015} and new demyelinating lesions on follow-up imaging may require modification to immunotherapy~\cite{McNamara2017}.
Change detection requires a balance between highlighting relevant differences, and suppressing trivial perceptual differences. The latter may be related to imperfect image registration or technical differences in acquisition, such as different scanners, different magnetic field strength (e.g. 1.5T versus 3T MRI), and different imaging spatial resolution.


Deep learning models have gained significant attention in recent years due to their unprecedented performance in a variety of computer vision tasks~\cite{Russakovsky2015}. Many of these have been translated to medical imaging applications~\cite{Lundervold2019}, including supervised lesion detection and classification~\cite{Huang2017}, organ and structural segmentation~\cite{Kayalibay2017,snaauw2019end}, and image enhancement~\cite{Plassard2018}. 
A drawback of supervised deep learning models is the requirement for a sufficiently large and balanced annotated training set~\cite{Hofmanninger2020}. 
Unfortunately, certain types of lesion appearance and change detection, such as MS in brain MRI, are anomalous events, where a large variety of changes can occur.
For such problems, it is prohibitively inefficient to annotate every type of change to collect a large and balanced training set. 
Thus, supervised models for change detection may not generalise well for broad application in medical imaging. 

In this paper, we propose a solution to mitigate the lack of annotated lesion change samples with a new unsupervised anomaly detection and localisation method trained only with serial images that do not contain any change.
Our training simulates changes in the serial images, enabling the production of detection and localisation pseudo-labels that are used to self-supervise the training of our model.
The images are synthesised by mixing super-pixels~\cite{Achanta2012} from the original image and its reconstructed image generated by a variational auto-encoder~\cite{Kingma2013}.
To tackle the class imbalance problem, we employ a focal Tversky loss~\cite{Abraham2018,Salehi2017}. 
Experiments on the problem of detection and localisation of new demyelinating lesions on longitudinal MRI in multiple sclerosis patients show that our method produces competitive detection and localisation results compared with other fully-supervised methods~\cite{Denner2020,kruger2020fully} trained on different datasets.

\section{Related Work}

Automated change detection is a long-standing problem in medical imaging~\cite{bosc2003automatic} and other fields~\cite{radke2005image,khelifi2020deep}.
Previous work on detecting change in longitudinal multiple sclerosis imaging 
include subtraction techniques to visualise areas of change~\cite{Patel2017}, statistical modelling~\cite{Schmidt2019}, and deep learning~\cite{Birenbaum2017,McKinley2020,kruger2020fully,Sepahvand2020}. Approaches that jointly solve image registration and change detection have also been devised~\cite{bu2020mask,dufresne2020joint}. 






We propose a data augmentation strategy based on mixing super-pixels~\cite{Achanta2012}.
Unlike CutOut~\cite{DeVries2017} and CutMix~\cite{Yun2019}, mixing based on super-pixels utilises a segmentation that is not restricted to horizontally- or vertically-oriented edges. Furthermore, as super-pixels group pixels with perceptually similar characteristics, mixing super-pixels is more likely to preserve the integrity of anatomical and lesional boundaries and maintain the saliency of transposed portions. 

\section{Materials and Methods}

\subsection{Data Acquisition and Processing}

Research ethics for this study was granted by Bellberry Limited. 
Longitudinal volumetric T2 FLAIR examinations performed by a private imaging provider (Anonymous Medical Imaging) between August 2018 and October 2020 across multiple sites were extracted. Based on findings in the radiologist report, pairs of scans were separated into two sets, namely: pairs containing lesion changes (Change), and pairs without lesion changes (NoChange). 
The training set contains pairs of scans in the NoChange set and is denoted by $\mathcal{D} = \{ \mathbf{x}_i,\hat{\mathbf{x}}_i \}_{i=1}^{|\mathcal{D}|}$, where $\mathbf{x}_i,\hat{\mathbf{x}}_i \in \mathcal{X} \subset \mathbb{R}^{H \times W \times D}$ represent a pair of the MRI scans from the same patient.
The testing set, with pairs of scans from the Change set, is represented by $\mathcal{T} = \{ (\mathbf{x}_i,\hat{\mathbf{x}}_i,\hat{\mathbf{y}}_i) \}_{i=1}^{|\mathcal{T}|}$, where  $\mathbf{x}_i,\hat{\mathbf{x}}_i \in \mathcal{X}$ denote a pair of scans from the same patient with a lesion change indicated in the binary map $\hat{\mathbf{y}}_i  \in \mathcal{Y} \subset \{0,1\}^{H \times W \times D}$.

The images in both sets were pre-processed with a resampling to 1.0mm $\times$ 1.0mm $\times$ 1.0mm isometric voxels, greyscale intensity normalisation to $[0, 1]$ (0th and 99th percentile), skull stripping ~\cite{Isensee2019}, rigid body registration~\cite{Garyfallidis2014}, and N4 bias correction~\cite{Lowekamp2013}.
For pairs of scans in the Change set, the ground truth masks for regions of change were manually annotated by a trainee radiologist (Anonymous) using ITK-SNAP~\cite{Yushkevich2006} based on the findings in the report as a guide. In total, 237 pairs of NoChange scans were used for model training, while 94 pairs of Change scans were used for model testing. 

\subsection{Methods}

The training of our model follows a two-stage process that uses only the NoChange dataset $\mathcal{D}$.  The first stage consists of a variational auto-encoder (VAE)~\cite{Kingma2013} combined with a new data augmentation strategy related to CutMix~\cite{Yun2019}, that automatically generates images and annotation maps with synthetic lesion changes. The second step comprises training of a siamese 3D U-net \cite{Ronneberger2015} that takes two serial images of the same patient and returns a segmentation map of the lesion change. 

\subsubsection{Generating Synthetic Lesion Changes} 
The generation of synthetic lesion changes starts with a VAE~\cite{Kingma2013} that relies on an encoder $f_{\theta}:\mathcal{X} \rightarrow \mathcal{M} \times \mathcal{S}$, a sampling from the Gaussian model to generate the embedding $\mathbf{z} \sim \mathcal{N}(\mu,\Sigma)$, where $\mathbf{z} \in \mathcal{Z}$, $\mu \in \mathcal{M}$ and $\Sigma \in \mathcal{S}$, and a reconstruction from the decoder $g_{\gamma}:\mathcal{Z} \rightarrow \mathcal{X}$ .  
This VAE is trained with the images $\mathbf{x} \in \mathcal{D}$ using an voxel-wise L1 reconstruction loss and a regularisation loss to minimise the Kulback-Leibler divergence between the embedding distribution and a zero-mean identity-covariance Gaussian.
To increase the diversity of the reconstructions, a perturbation is also applied to the embedding, $\tilde{\mathbf{z}} = \mathbf{z}\times \Delta$, where $\Delta \sim \mathcal{U}(-\delta,\delta)$ ($\mathcal{U}(.)$ denoting a uniform distribution).
Hence, for each image $\mathbf{x}_i \in \mathcal{D}$, we sample new reconstructions, denoted by $\tilde{\mathbf{x}}_i = g_{\gamma}(\mathbf{z}_i \times \Delta)$, with $\mathbf{z}_i \sim \mathcal{N}(\mu_i,\Sigma_i)$, and $\mu_i,\Sigma_i=f_{\theta}(\mathbf{x}_i)$.

\begin{figure}[b!]
\includegraphics[width=\textwidth]{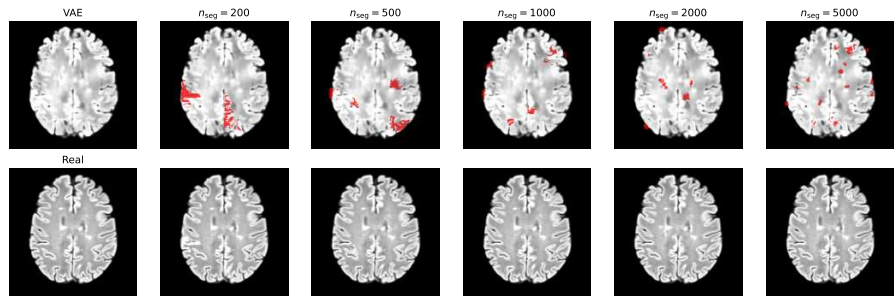}
\caption{SuperMix examples showing the effect of $n_\mathrm{seg}$. The perturbed VAE reconstruction ($\delta = 5$) and overlaid SuperMix masks (red) are shown in the top row. The resultant synthesised lesion changes are shown in the bottom row.} \label{fig:superMix}
\end{figure}

To synthesise lesion changes, we propose a super-pixel mixing data augmentation strategy, referred to as SuperMix.
Using the scan $\mathbf{x}_i$ and its sampled reconstruction $\tilde{\mathbf{x}}_i$, we first produce a super-pixel segmentation~\cite{Achanta2012} that tesselates $\mathbf{x}_i$ to produce $n_{\text{seg}}$ binary maps, each represented by 
$\mathbf{b}_{i,t} \in \{0,1\}^{H \times W \times D}$ for $t \in \{0,...,n_{\text{seg}}-1\}$ (in this map, a voxel is labelled with 1, if it belongs to the $t^{th}$ super-pixel). The synthesised image and lesion change are defined by
\begin{equation}
\begin{split}
    \mathbf{x}^\prime_i &= \sum_{t=0}^{n_{\text{seg}}-1} (\lambda_t) (\mathbf{b}_{i,t} \odot \mathbf{x}_i) + (1-\lambda_t)(\mathbf{b}_{i,t} \odot \tilde{\mathbf{x}}_i),\\
    \hat{\mathbf{y}}_i &= \sum_{t=0}^{n_{\text{seg}}-1} 
    (\lambda_t)(\mathbf{0}_{H\times W \times D}) + 
    (1-\lambda_t)(\mathbf{b}_{i,t}),
\end{split}
\label{eq:synth_image}
\end{equation}
where $\lambda_t = 1$ if $u < \tau$, with $u \sim \mathcal{U}(0,1)$, and $\mathbf{0}_{H\times W \times D}$ denotes a volume of size $H\times W \times D$ containing only zeros. This allows us to build a synthesised set of Change, represented by $\hat{\mathcal{D}} = \{ (\mathbf{x}^\prime_i,\hat{\mathbf{x}}_i,\hat{\mathbf{y}}_i) \}_{i=1}^{|\mathcal{D}|}$. Fig.~\ref{fig:superMix} shows examples of the synthesised lesion changes for different values of $n_{\text{seg}}$.


\subsubsection{Siamese 3D U-net}
\begin{figure}[t!]
	\centering
	\includegraphics[width=.90\textwidth]{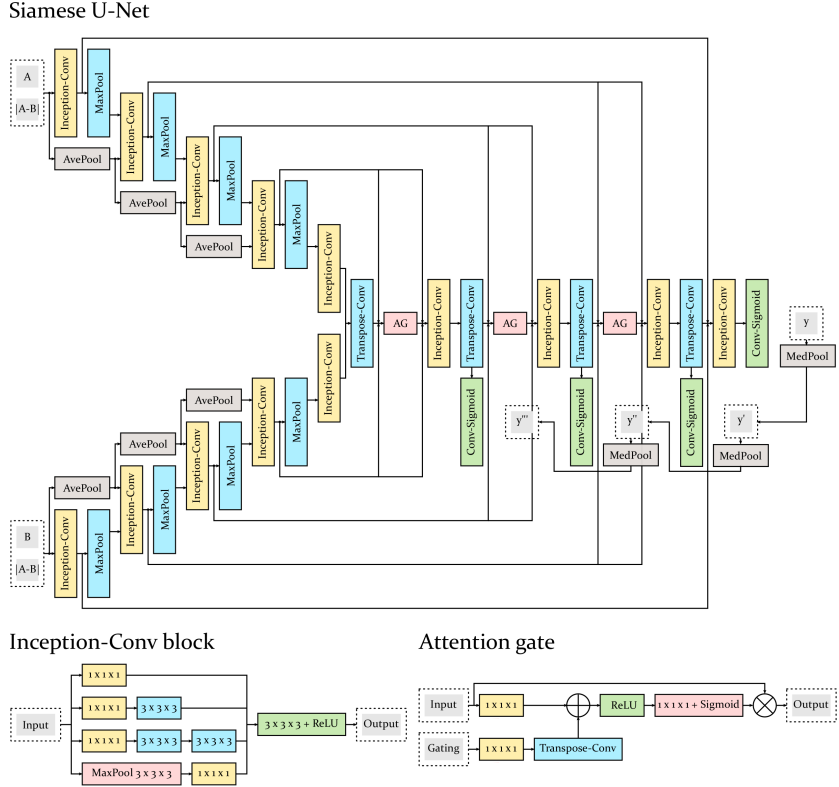}
	\caption{Schematic of proposed siamese U-net model.}
	\label{fig:discriminator}
\end{figure}

Our proposed siamese 3D U-net takes a sample $(\mathbf{x}^\prime_i,\hat{\mathbf{x}}_i,\hat{\mathbf{y}}_i) \in \hat{\mathcal{D}}$ 
to form two inputs and one output.  
The inputs are $\mathbf{x}_i^{(1)}=(\mathbf{x}^\prime_i,|\mathbf{x}^\prime_i - \hat{\mathbf{x}}_i|) \in \mathcal{X}^2$ and
$\mathbf{x}_i^{(2)}=(\hat{\mathbf{x}}_i,|\mathbf{x}^\prime_i - \hat{\mathbf{x}}_i|)  \in \mathcal{X}^2$, and the output is the segmentation map $\hat{\mathbf{y}}_i$.
Siamese network configurations enable independent processing of image inputs that may improve feature extraction prior to fusion and comparison of features~\cite{Daudt2018}.
Our siamese 3D U-net consists of a down-sampling network, represented by $e_{\phi}:\mathcal{X}^2 \times \mathcal{X}^2 \rightarrow \mathcal{K}$, 
followed by an up-sampling network, denoted by $d_{\psi}:\mathcal{K} \rightarrow [0,1]^{H \times W \times D}$.
As shown in Fig.~\ref{fig:discriminator} the model contains skip connections that propagate data directly from the encoder to decoder at the same spatial resolution, providing local contextual information to the up-sampling process.
Each block of the model is formed by inception modules~\cite{Szegedy2015} that incorporate convolutional filters of different sizes with dimensionality reduction achieved by $1 \times 1 \times 1$ convolutional layers to enable efficient processing of features at multiple scales. 
Since areas of change typically occupy small regions in the entire image, the ability to focus the network to only relevant areas improves utilisation of the network resources. To that end, attention gates were built into the network to filter information being passed through the skip connections~\cite{Schlemper2019}.
Down-sampling operations in the U-net encoder can result in loss of information flow in the deeper layers. To mitigate this, we provided down-sampled inputs to the deep encoding layers. We also encouraged the intermediate layers of the U-net decoder to learn meaningful segmentations by incorporating deep supervision~\cite{Lee2014}. This was implemented by passing the outputs of the intermediate layers through a $1 \times 1 \times 1$ convolutional layer with sigmoid activation.


The training of this siamese 3-D model is based on the minimisation of the focal Tversky loss~\cite{Abraham2018,Salehi2017}  that can handle the  class imbalance problem associated with segmenting small and localised areas of change.  This loss is defined by
\begin{equation}
    \ell(\hat{\mathcal{D}},\phi,\psi) = \frac{1}{|\mathcal{D}|}\sum_{i=1}^{|\mathcal{D}|}(1 - \ell_{tv}(\hat{\mathbf{y}}_i,\tilde{\mathbf{y}}_i))^{\gamma},
\label{eq:focalTL}
\end{equation}
where $\gamma \in \mathbb{R}^{+}$,
$\tilde{\mathbf{y}}_i = d_{\psi}(e_{\phi}(\mathbf{x}_i^{(1)},\mathbf{x}_i^{(2)})))$, and
$\ell_{tv}(.)$, the Tversky index, generalises the Dice coefficient, with
\begin{equation}
    \ell_{tv}(\hat{\mathbf{y}}_i,\tilde{\mathbf{y}}_i) = \frac{TP}{TP + \alpha FN + \beta FP},
\label{eq:TL}
\end{equation}
where $TP$, $FN$ and $FP$ are the numbers of true positive, false negative and false positive between the annotation $\hat{\mathbf{y}}_i$ and the model output $\tilde{\mathbf{y}}_i$, respectively. 
In~\eqref{eq:TL}, the parameters $\alpha$ and $\beta$ can be tuned to emphasise recall over precision ($\alpha > \beta$), or vice versa ($\alpha < \beta$), such that $\alpha + \beta = 1$ (Supplemental Fig. 3).

After training the model, inference is performed by running it on a test sample composed of a pair of serial images $(\mathbf{x},\hat{\mathbf{x}})$ from $\mathcal{T}$.  The model produces $\tilde{\mathbf{y}} = d_{\psi}(e_{\phi}(\mathbf{x}^{(1)},\mathbf{x}^{(2)}))$ with $\mathbf{x}^{(1)}=(\mathbf{x},|\mathbf{x} - \hat{\mathbf{x}}|)$ and
$\mathbf{x}^{(2)}=(\hat{\mathbf{x}},|\mathbf{x} - \hat{\mathbf{x}}|)$.
Then, we threshold this output to produce a binary output with
$\tilde{\mathbf{y}} > \kappa$, which is then used by connected component analysis (CCA) to produce segmentation blobs of size $\ge \{20,100\}$.
This binarised result can then be compared with the ground truth annotation in $\hat{\mathbf{y}}$ from $\mathcal{T}$.

\section{Experiments and Evaluation}
Training for the VAE consisted of 200 iterations in batches of 4096 volume crops randomly sampled from the set of NoChange scans, with mini batch size of 2. Each crop measured 192 $\times$ 192 $\times$ 16 voxels. The effect of $\delta$ to paramaterise the uniform distribution that influences the VAE generation is linked to the VAE encoder architecture, such that replacing max-pooling with strided convolutions results in greater mottling in the generator output (Supplemental Fig. 1). For all experiments we utilised the VAE (MaxPool) architecture and $\delta$ was fixed at 5.0.
The training for the lesion change detector consisted of 60 iterations in batches of 100 NoChange pairs augmented by SuperMix, with mini batch size of 2. 
The SuperMix mixing parameter $\tau$ in~\eqref{eq:synth_image} was fixed at 0.98, with $\log{n_\mathrm{seg} \sim \mathcal{U}(\log{200},\log{5000})}$. 
The recall/precision balance of the model is tunable by the $\alpha$, $\beta$, and $\gamma$ parameters of the focal loss in~\eqref{eq:focalTL} and \eqref{eq:TL} (Supplemental Fig. 3). For all outputs $\alpha$ was set to 0.75, while $\gamma=0.75$ for intermediate layers, and we tested $\gamma = \{0.75,1.0\}$ for the final layer.  
In a series of ablation experiments, we explored the contributions of network architecture (non-siamese or siamese U-net), loss function (binary cross-entropy or focal Tversky loss), multi-scale inputs, deep supervision, attention gates, and Inception modules.  We included L2 weight regularisation in the siamese U-net. 

Adam optimiser was used. A learning rate of 0.00005 was used to train the VAE. For the lesion change detector, an initial learning rate of 0.0002 and learning decay rate of 0.001 were used. Models were written in Keras/Tensorflow and trained on a GeForce RTX 3090 GPU with 24GB of memory. 

Model performance was evaluated by generating prediction masks on the Change pairs, thresholding $\kappa = 0.1$, and comparing against ground truth lesions. 
The lesion-wise true positive rate (LTPR), lesion-wise false positive rate (LFPR) and positive predictive value (PPV)
were calculated for lesion changes in each Change pair and averaged across all pairs (Supplemental D). 
Overlap between a ground truth change and predicted change lesion was significant (i.e. a lesion-wise true positive) if the intersection over union (IoU) of the blobs was 0.01 or greater. Others have considered overlap of a single voxel to be sufficient~\cite{Aslani2019}, but we decided to be more conservative with an IoU$\ge .01$. 



\setlength{\tabcolsep}{0.5em} 

\subsection{Model performance}

\begin{figure}[t!]
\centering
\includegraphics[width=0.83\textwidth]{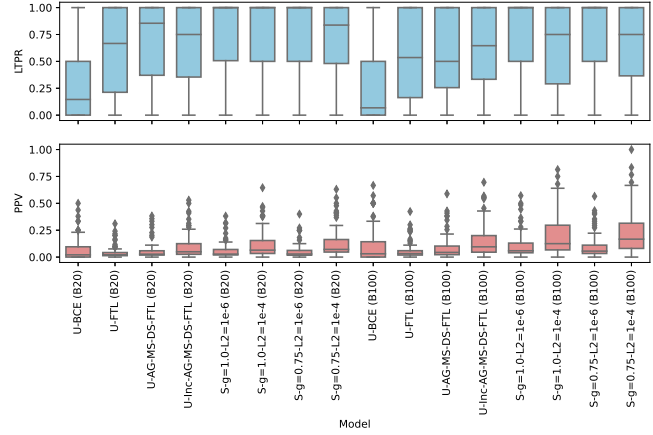}
\vspace{-8mm}
\caption{Performance of different model configurations. U, U-net, BCE, binary cross-entropy; FTL, focal Tversky loss; AG, attention gate; MS, multiscale input; DS, deep supervision; B, minimum blob size; g, $\gamma$ parameter of final layer FTL; L2, L2 weight regularisation factor.} 
\label{fig:boxplots}
\end{figure}

\begin{figure}[b!]
\centering
	\begin{subfigure}[b]{0.28\textwidth}
	\centering
    \includegraphics[width=\textwidth]{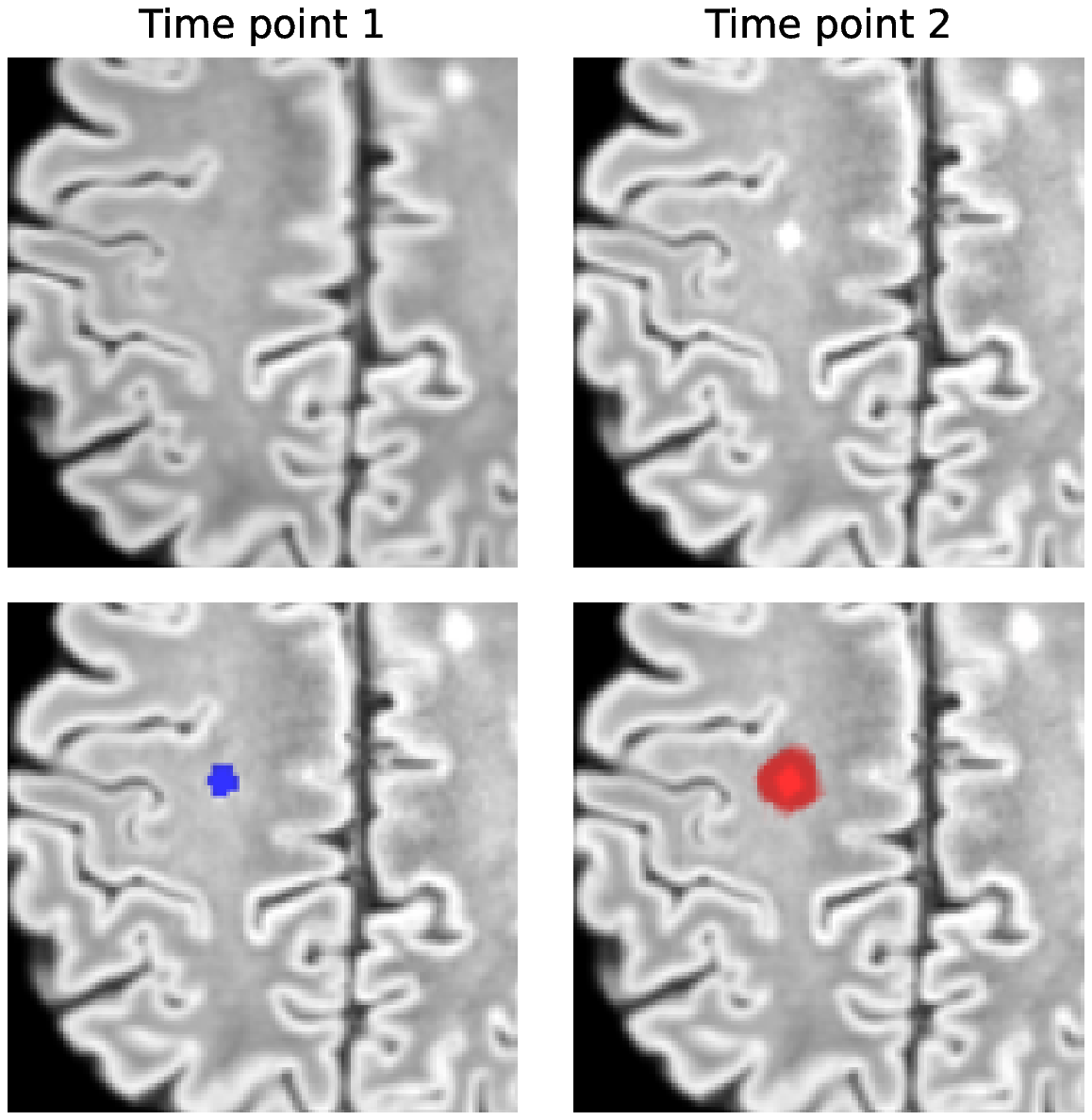}	
    \caption{Cortical}
	\end{subfigure}
	\hfill
	\begin{subfigure}[b]{0.28\textwidth}
	\centering
	\includegraphics[width=\textwidth]{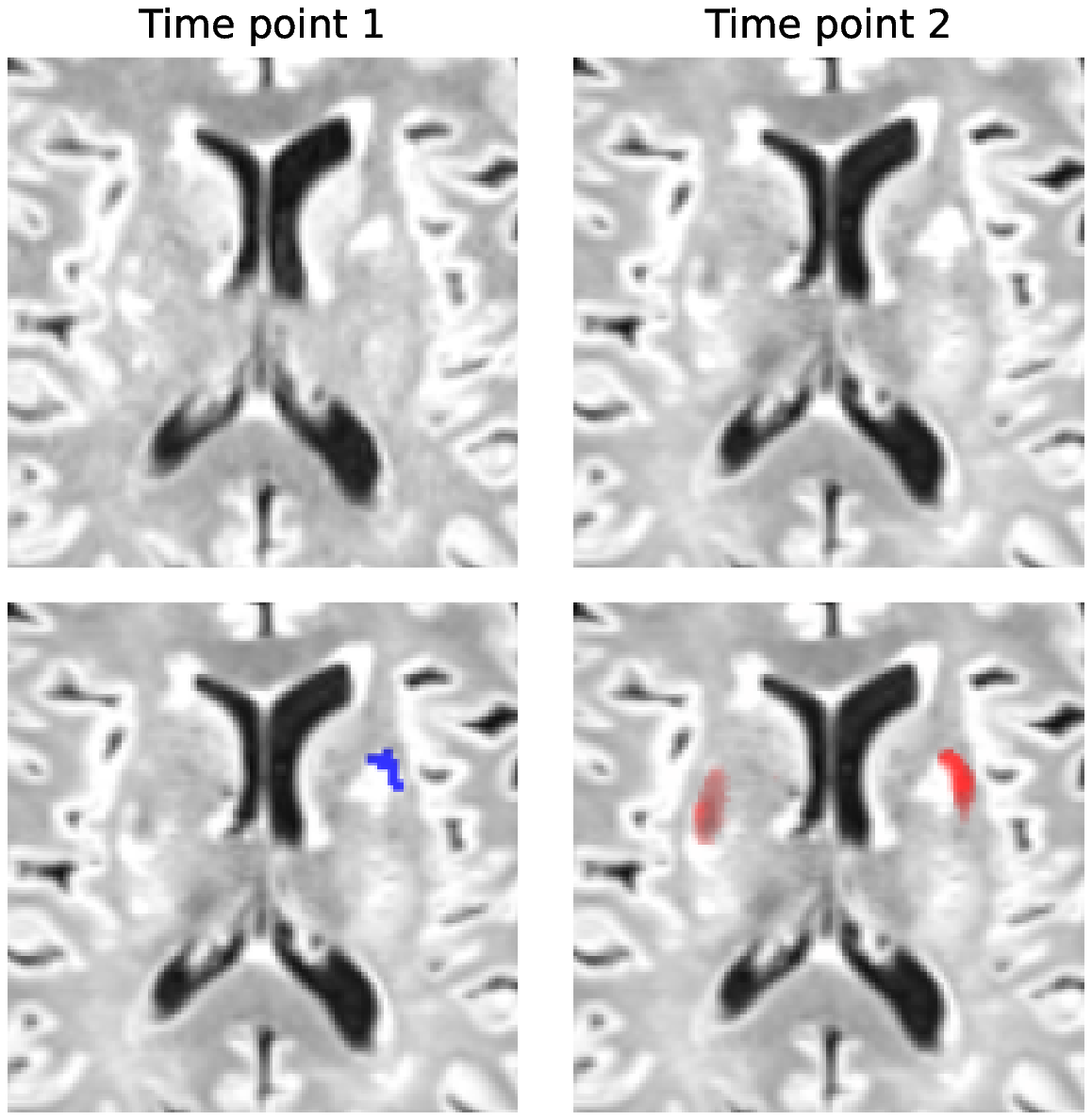}
	\caption{Subcortical}
	\end{subfigure}
	\hfill
	\begin{subfigure}[b]{0.28\textwidth}
	\centering
	\includegraphics[width=\textwidth]{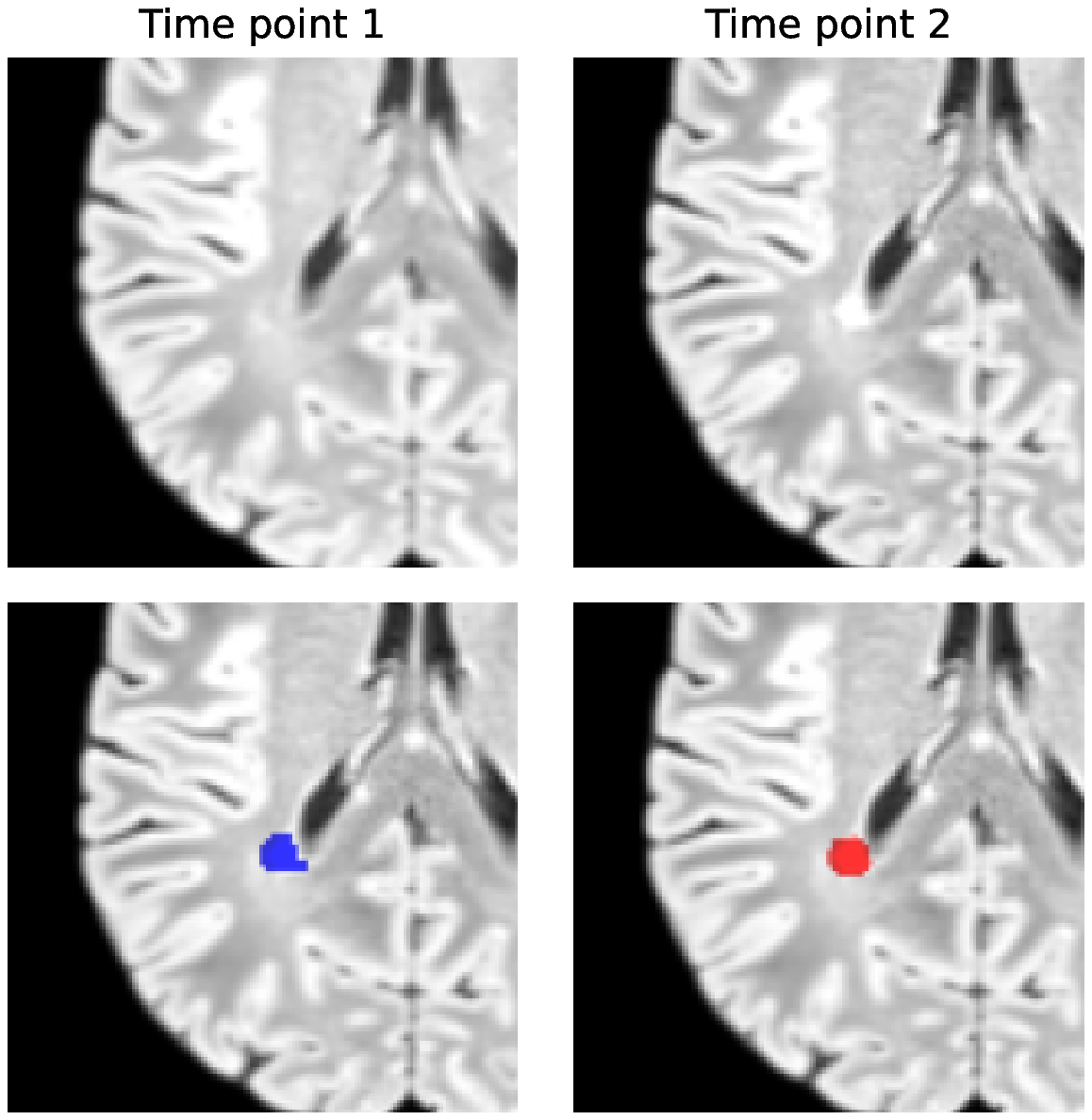}
	\caption{Peri-ventricular}
	\end{subfigure}
	
	\begin{subfigure}[b]{0.28\textwidth}
	\centering
	\includegraphics[width=\textwidth]{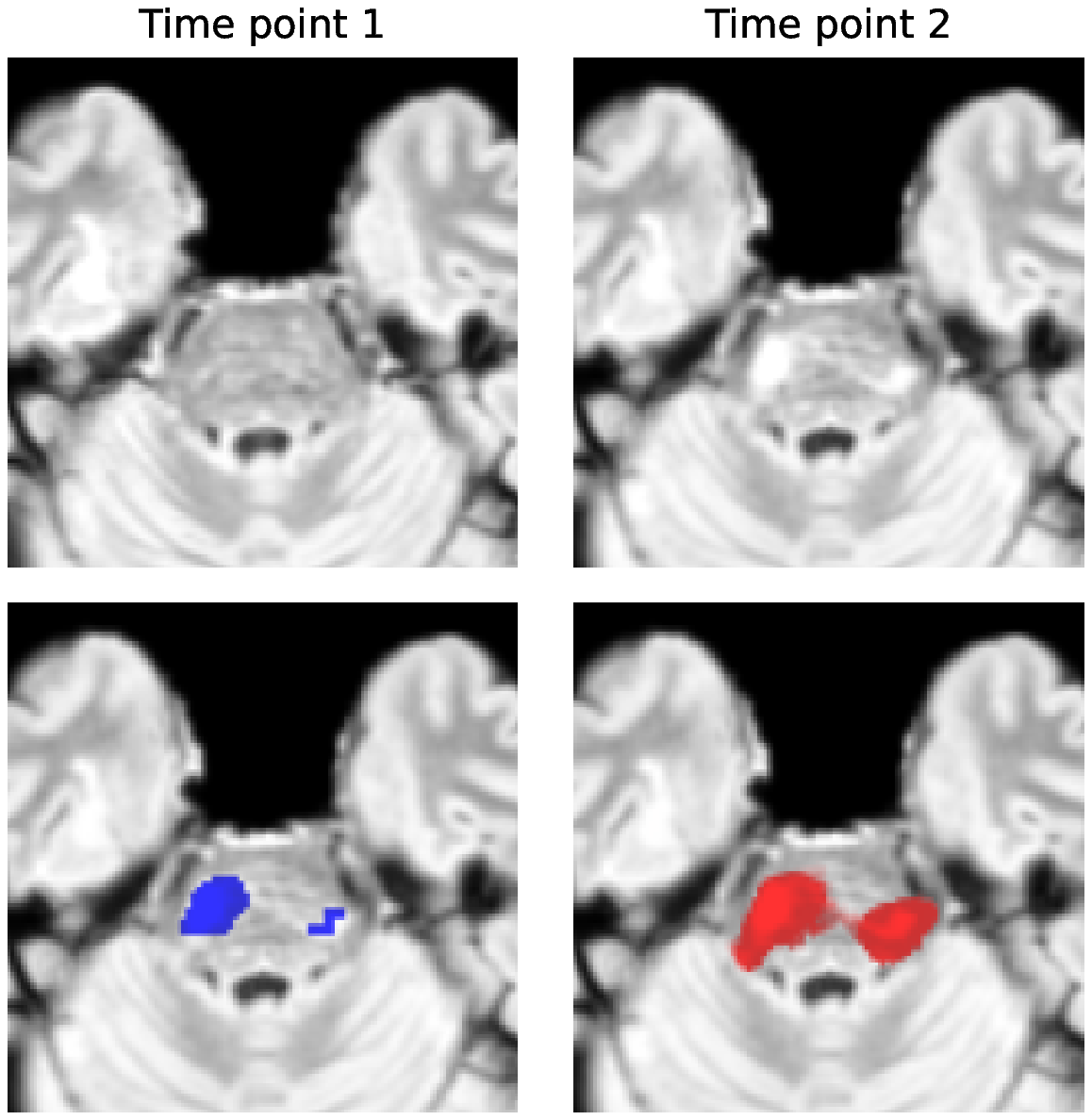}
	\caption{Brainstem}
	\end{subfigure}	
	\hfill
	\begin{subfigure}[b]{0.28\textwidth}
	\centering
	\includegraphics[width=\textwidth]{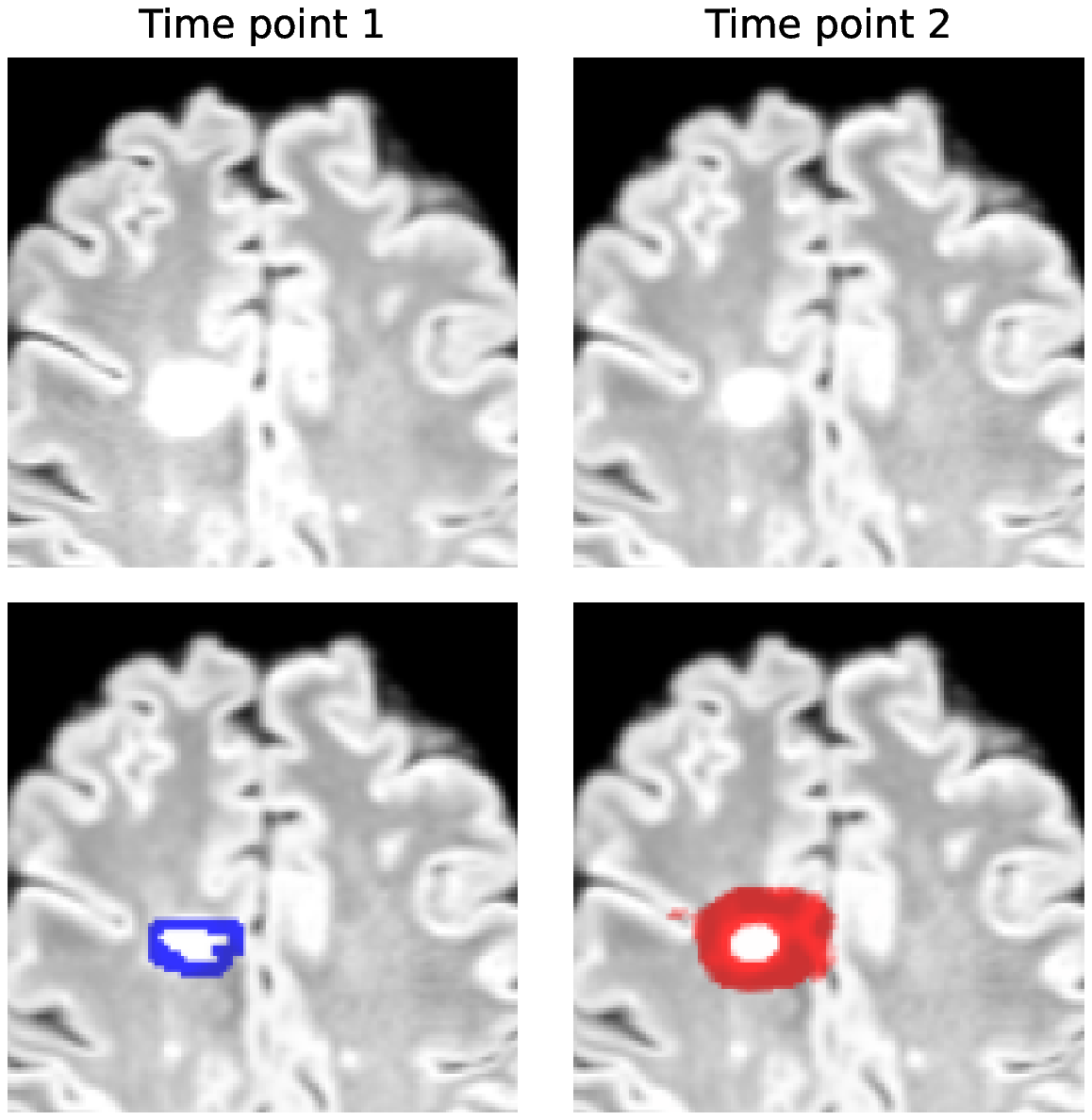}
	\caption{Lesion expansion}
	\end{subfigure}	
	\hfill
	\begin{subfigure}[b]{0.28\textwidth}
	\centering
	\includegraphics[width=\textwidth]{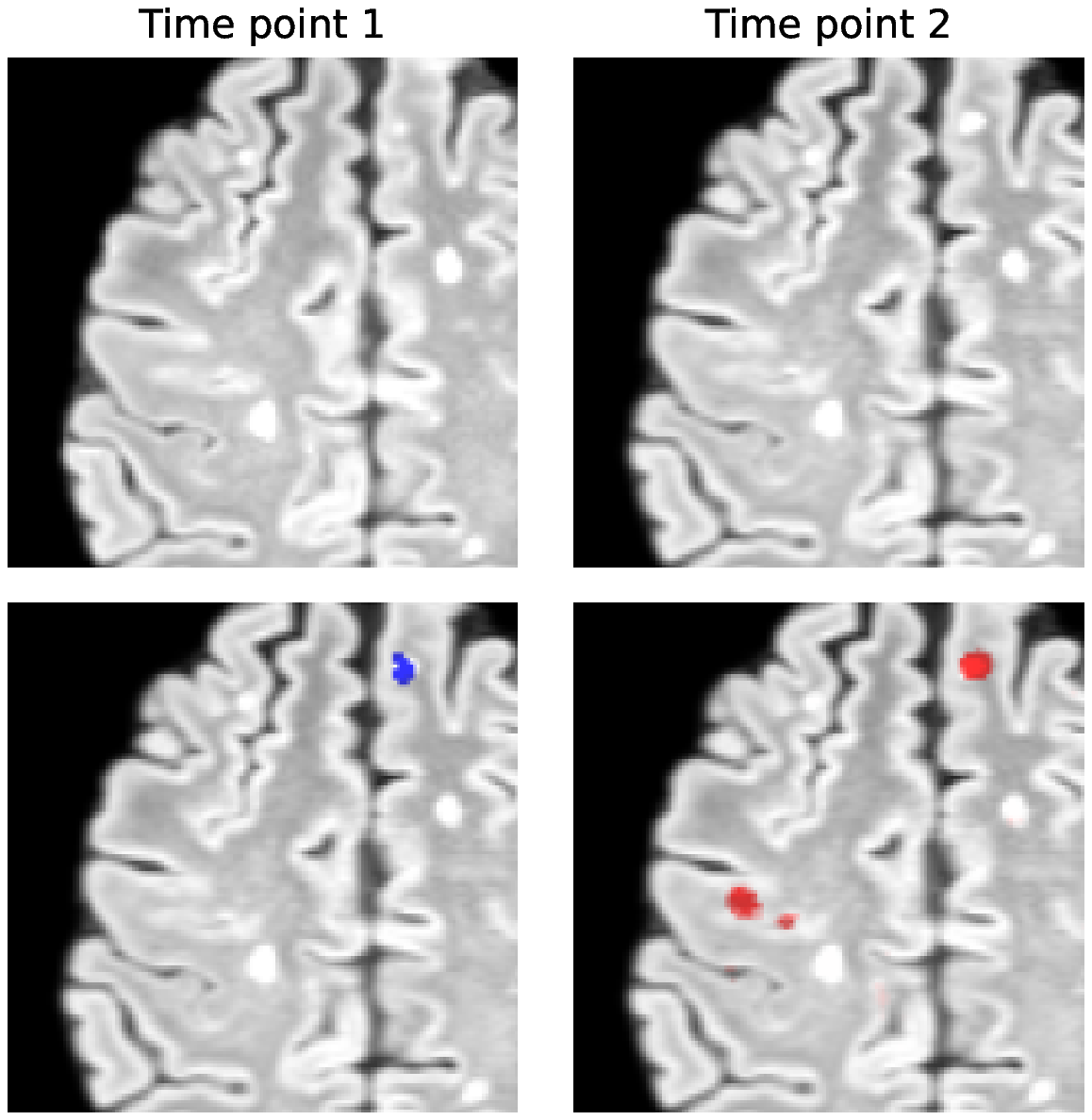}
	\caption{False detection}
	\end{subfigure}
\caption{Example predicted change masks. Ground truth (blue) and predicted (red) change masks are shown overlaid the scan at time point 2.} 
\label{fig:modelpred}
\end{figure}

Ablation experiments show the incremental improvements in change detection provided by incorporation of the focal Tversky loss, attention gates, multiscale inputs and deep supervision (Fig. \ref{fig:boxplots} and Table \ref{tab:results}). Connected components post-processing with a larger blob size (100 compared to 20) generally improve PPV, with a corresponding reduction in LTPR. 
Siamese architectures yielded the best performance. The focal loss parameter $\gamma$ did not strongly influence segmentation performance. 
Example change mask predictions are shown in Fig. \ref{fig:modelpred}, demonstrating the model's ability to detect changes in different parts of the brain, as well as changes to existing lesions. 
High LFPR values may correspond to other regions of change (Fig. \ref{fig:modelpred}(f)), distinct from lesion changes. Some may be changes unrelated to multiple sclerosis, but their relevance is not explored here.

The lesion change recall (i.e. LTPR) of the siamese models is comparable to published supervised models \cite{Denner2020,kruger2020fully}. We also test a publicly-available deep fully convolutional network model for white matter hyperintensities (WMH) segmentation that was trained on different datasets~\cite{Li2018}. In this case, the difference between the predicted WMH segmentations of each input image is taken to be the predicted change segmentation. Despite reporting an average recall of 0.84 on the original data, that model~\cite{Li2018} performs poorly on our data. We observed that lesions tended to be under-segmented and lesion changes were frequently missed, resulting in a low LTPR the model in~\cite{Li2018} (Table \ref{tab:results}).

\begin{table}[t!]
\caption{Performance of selected model configurations. AG, attention gate; MS, multiscale input; DS, deep supervion; BCE, binary cross-entropy; FTL, focal Tversky loss. See also Fig.~\ref{fig:boxplots}.}\label{tab:results} 
\centering
\scalebox{0.8}{
\begin{tabular}{lccccccc}

\hline
Model & Loss & $\gamma$ & L2 reg   & Blob & LTPR & LFPR & PPV \\
\hline
\hline
U-net               & BCE   & -         & -     &   20  & 0.330 & 0.932 & 0.0684\\
U-net               & FTL   & 1.0 & -           &   20  & 0.618 & 0.960 & 0.0404\\
U-net-AG-MS-DS      & FTL   & 1.0 & -           &   20  & 0.664 & 0.945 & 0.0546\\
U-net-Inc-AG-MS-DS  & FTL   & 1.0 & -           &   20  & 0.645 & 0.909 & 0.0914\\
\hline
Siamese U-net       & FTL   & 1.0 & $10^{-6}$   &   20  & \textbf{0.758} & 0.942 & 0.0579\\
Siamese U-net       & FTL   & 1.0 & $10^{-4}$   &   20  & 0.708 & 0.884 & 0.116\\
Siamese U-net       & FTL   & 0.75 & $10^{-6}$  &   20  & 0.747 & 0.946 & 0.0541\\
Siamese U-net       & FTL   & 0.75  & $10^{-4}$ &   20  & 0.634 & \textbf{0.874} & \textbf{0.126}\\
\hline
Li et al. 2018~\cite{Li2018}             & -     & -         & -   &   -  & 0.133 & 0.997     & 0.0031 \\
\hline
\end{tabular}}
\vspace{-5mm}
\end{table}


\section{Conclusions}
The generation of images and annotation maps with synthetic lesion changes and the siamese U-net are the main contributions of this paper.
These contributions mitigate the  lack of annotated lesion change datasets with a self-supervised training of the anomaly detection and localisation methods. 
Change detection is optimised by the incorporation of the focal Tversky loss, attention gates, multiscale inputs and deep supervision.
Our approach, which does not require detailed, voxel-level annotated training sets, demonstrates high detection rates for lesion changes in multiple sclerosis imaging, comparable to fully-supervised models. 


%
%
%

\newpage
\bibliographystyle{splncs04}
\bibliography{references}

\end{document}


\title{Supplemental Material}

\titlerunning{Change detection in multiple sclerosis imaging}

\author{}

\institute{}

\maketitle              

\setstretch{1.0}

\section{VAE perturbation}
\begin{figure}
	\centering
    
    \begin{subfigure}[b]{\textwidth}
        \centering
        \includegraphics[scale=0.6]{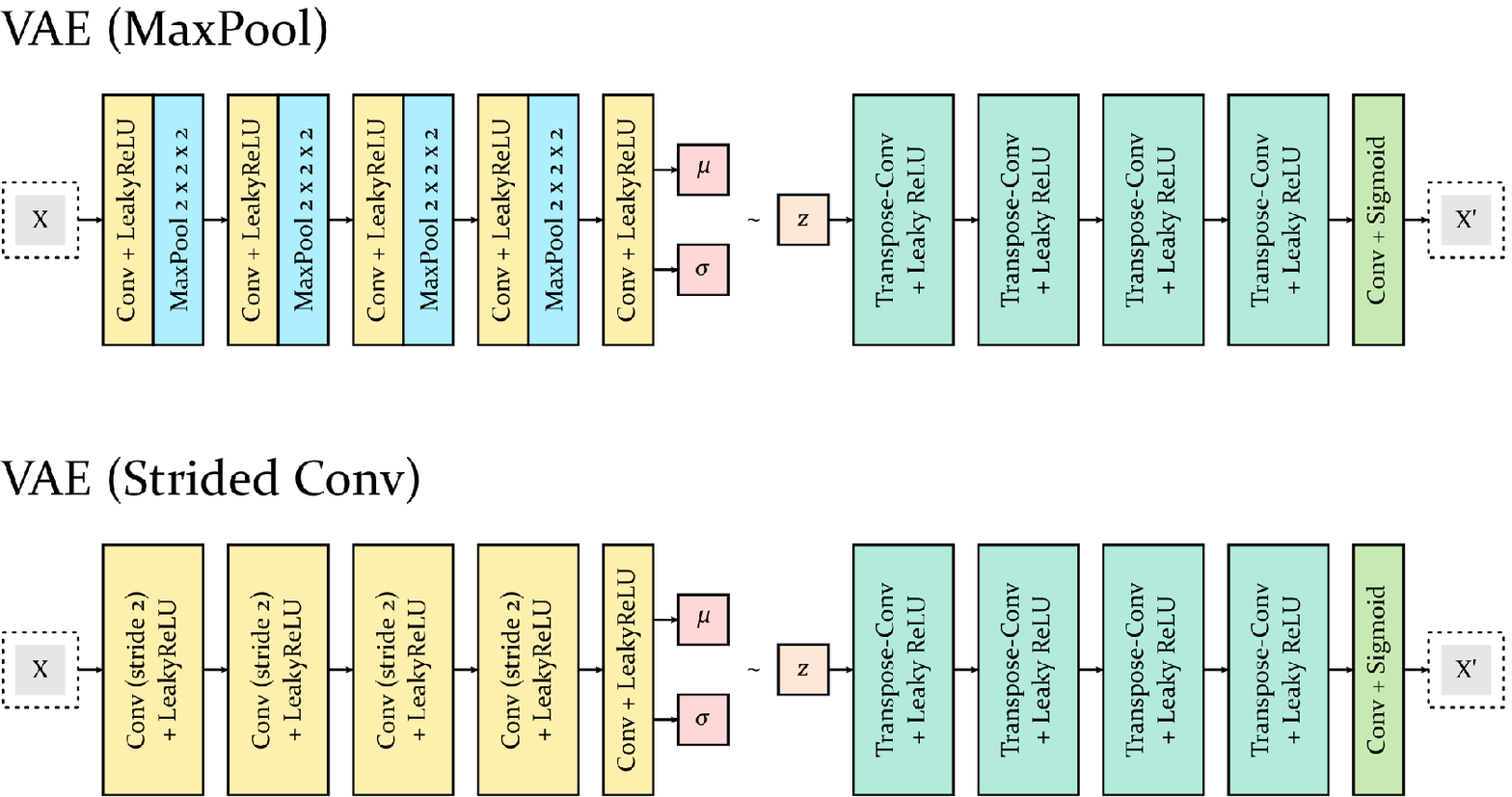}
        \caption{VAE architectures}
        \label{suppfig:vaearch}
    \end{subfigure}
    \hfill
    \begin{subfigure}[b]{\textwidth}
        \centering
        \includegraphics[width=\textwidth]{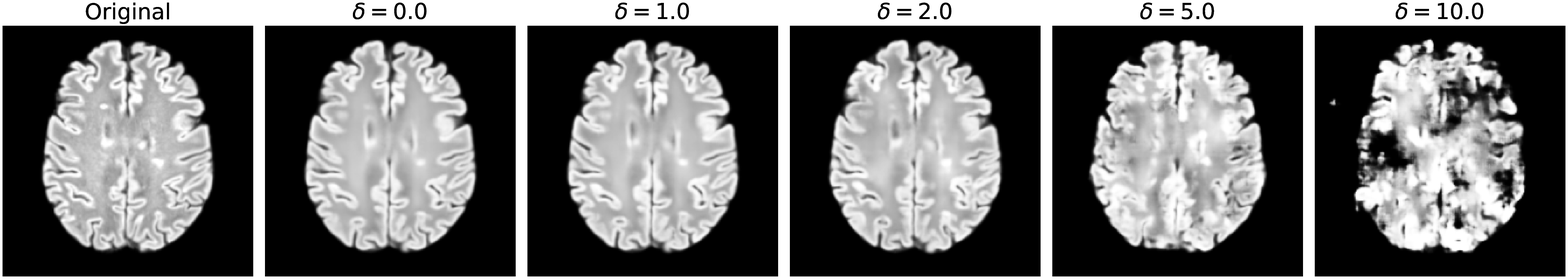}
        \includegraphics[width=\textwidth]{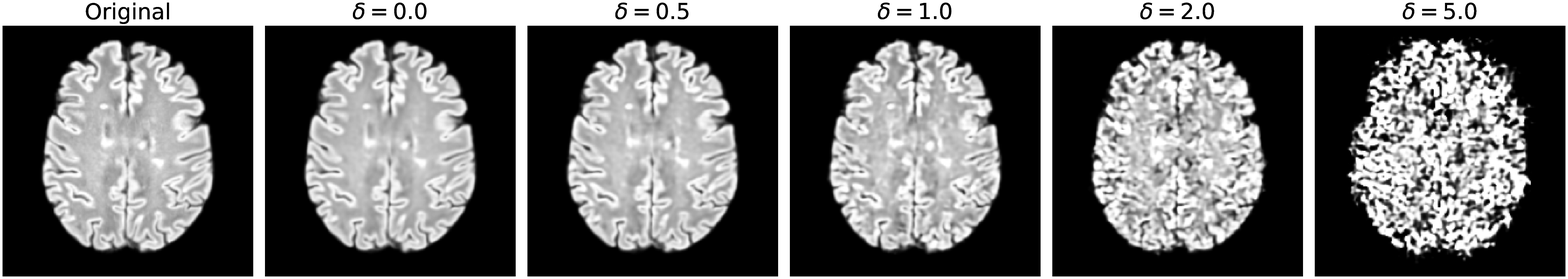}
        \caption{Corresponding images produced by VAE peturbation}
        \label{suppfig:vaeout}
    \end{subfigure}

	\caption{Variational auto-encoder architectures used to generate perturbations of input image.}
	\label{suppfig:vae}
\end{figure}

\newpage
\section{Deep supervision}
\vspace{-5mm}
\begin{figure}[h!]
	\centering
        \includegraphics[width=\textwidth]{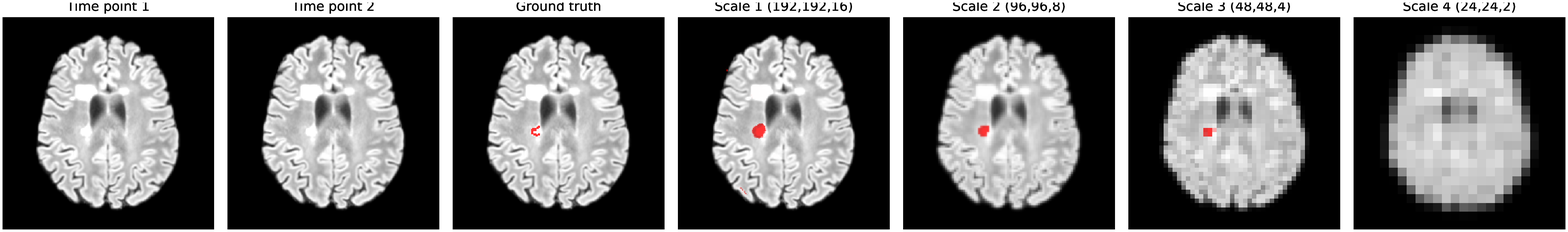}
        \includegraphics[width=\textwidth]{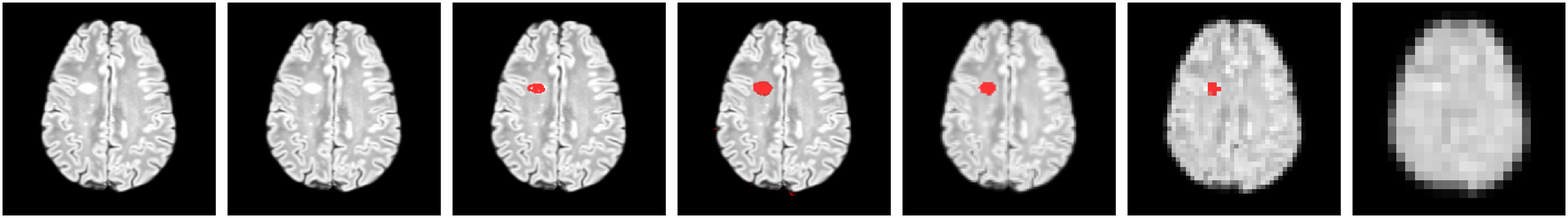}
	\caption{Intermediate predictions with deep supervision.}
	\label{suppfig:deepsuper}
\end{figure}

\section{Tuning recall and precision}
\vspace{-5mm}
\begin{figure}[h!]
	\centering
        \includegraphics[width=\textwidth]{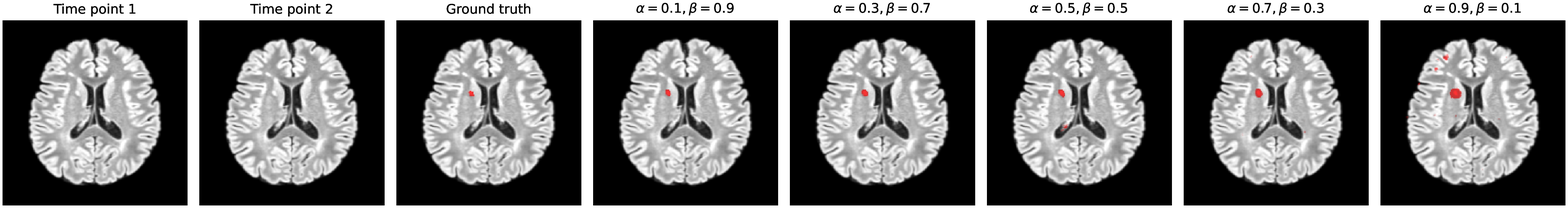}
        \includegraphics[width=\textwidth]{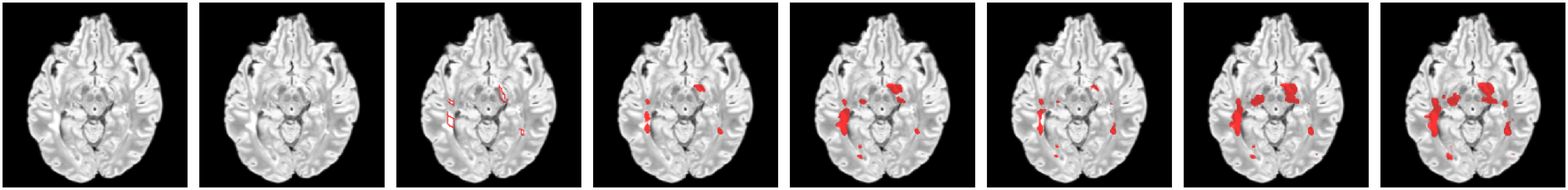}
	\caption{The effect of tuning $\alpha$ and $\beta$ parameters of the focal Tversky loss.}
	\label{suppfig:tuning}
\end{figure}

\section{Evaluation metrics}

The lesion-wise true positive rate (LTPR), lesion-wise false positive rate (LFPR) and positive predictive value (PPV) are defined as
\begin{eqnarray*}
    LTPR &=& \frac{LTP}{LTP + LFN}, \\
    LFPR &=& \frac{LFP}{LTP + LFP}, \\
    PPV &=& \frac{LTP}{LTP + LFP},
\end{eqnarray*}
where $LTP$, $LFN$, and $LFP$ denote lesion-wise true positives, false negatives, and false positives, respectively.